\documentclass[conference]{IEEEtran}

\ifCLASSINFOpdf
   \usepackage[pdftex]{graphicx}
   \graphicspath{{./figs/}}
     \DeclareGraphicsExtensions{.eps}
\else
  \usepackage[dvips]{graphicx}
   \graphicspath{{./figs/}}
   \DeclareGraphicsExtensions{.eps}
\fi
\usepackage[cmex10]{amsmath}
\usepackage{graphicx}
\usepackage{amsmath,amsthm,paralist,bm}
\usepackage{amssymb}
\usepackage{epstopdf}
\usepackage[sort&compress, numbers]{natbib}
\usepackage[usenames]{color}
\begin{document}
\IEEEoverridecommandlockouts 

\title{Compressive sensing for dynamic spectrum access networks: Techniques and tradeoffs}

\author{

\IEEEauthorblockN{
\it{
J.~N.~Laska,\IEEEauthorrefmark{1}
W.~F.~Bradley,\IEEEauthorrefmark{2}
T.~W.~Rondeau,\IEEEauthorrefmark{3}
K.~E.~Nolan,\IEEEauthorrefmark{4}
and B.~Vigoda\IEEEauthorrefmark{2}}
}\\
%
\IEEEauthorblockA{\IEEEauthorrefmark{1}
Dept. Electrical and Computer Engineering, Rice University, Houston, Texas, USA}
%
\IEEEauthorblockA{\IEEEauthorrefmark{2} Lyric Semiconductor,
  Cambridge, Massachusetts, USA} 
%
\IEEEauthorblockA{\IEEEauthorrefmark{3}
 GNU Radio, Philadelphia, Pennsylvania, USA}
%
\IEEEauthorblockA{\IEEEauthorrefmark{4}
 The Telecommunications Research Centre (CTVR), University of Dublin, Trinity College, Ireland}
\thanks{This research is partially sponsored by DARPA under Grant FA8750-07-C-0231.  It is also partially based upon work supported by Science Foundation Ireland under Grant No. 08/CE/I523 as part of CTVR / The Telecommunications Research Centre at University of Dublin, Trinity College, Ireland.  }
}

\maketitle

\begin{abstract}
We explore the practical costs and benefits of CS for
dynamic spectrum access (DSA) networks. Firstly, we review several fast 
and practical techniques for energy detection without full reconstruction 
and provide theoretical guarantees. We also define practical metrics to measure
the performance of these techniques. Secondly, we perform comprehensive
experiments comparing the techniques on real signals
captured over the air. Our results show that
we can significantly compressively acquire the signal while still accurately
determining spectral occupancy.

\end{abstract}

\IEEEpeerreviewmaketitle

\section{Introduction}
\label{sec:intro}

In most regions,  the \emph{radio frequency} (RF) spectrum is completely allocated; however, at any given time, it is likely to be severely underutilized~\cite{mchenry:xg:2007,marshall:dyspan:2008,Akylidiz06}. With recent advances in the size and power of digital processors, it has become possible to develop radios that can make dynamic transmission decisions.  Thus,  \emph{dynamic spectrum access} (DSA) networks have emerged with the primary goal of harnessing this additional spectrum to support an increasing number of wireless services. 


A key challenge in DSA networks is designing schemes that take
advantage of the unused spectrum capacity without communicating or
interfering with active transmissions.  There are two approaches
currently in practice.  The first approach, known as \emph{spectrum
  sensing}, monitors a wide band of RF spectrum and detects channels
of significant energy.  The second approach monitors a database that
tracks spectral usage based on channel and geographical area.  In both
cases, once the active channels are recognized, the DSA system can
then transmit into another available channel.  While interesting
problems arise from both techniques, in this paper we focus on the
former.



There are two fundamental bottlenecks restricting efficient spectrum sensing: 
\begin{enumerate}
\item the cost and fidelity of wideband analog-to-digital converters
  (ADCs); and
\item the power consumption and processing speed of the sensing
  system.
\end{enumerate}
The first bottleneck follows from the Nyquist-Shannon sampling theorem; an
arbitrary bandlimited signal must be acquired at a sample rate of
twice its bandwidth. It follows, then, that attempting to capture and
monitor large bandwidths requires very high speed ADCs. We also know
that the price and power consumption of an ADC is directly related to
its sampling rate \cite{le_adc2005}. Therefore, ADCs capable of
wideband spectrum sensing, along with the required RF front-end, are
expensive and consume prohibitive levels of power.  Furthermore, large
sampling rates result in the second bottleneck; a large number of
samples need to be processed. This will also negatively effect the
speed and power required to process the samples.

The recently developed \emph{compressive sensing} framework demonstrates that
signals can be recovered with significantly fewer samples than demanded
by Nyquist-Shannon sampling
theory~\cite{Can::2006::Compressive-sampling,Don::2006::Compressed-sensing,CanRomTao::2006::Stable-signal}. The
key insight is that we can exploit known signal structure beyond
simple bandlimitedness.  Specifically, CS enables the acquisition of
\emph{sparse} and approximately sparse signals. A sparse signal has
few non-zero coefficients in some transform domain. Similarly, an
approximately sparse signal has only a few large values and, instead
of zero coefficients, most of the samples are comparatively very small
(i.e., noise). CS requires a linear measurement system and non-linear
signal reconstruction algorithms; however, as we will discuss, for
some tasks, computation can be performed directly on the
samples~\cite{SPARS,DavDuaWak::2007::The-smashed-filter,DavBouWak::2010::Signal-processing,DavSchSla::2010::A-wideband-compressive}.

This radical framework has inspired several new sub-Nyquist
analog-to-digital converters, each aiming to acquire wide
bandwidth signals at lower sample rates than their Nyquist
counterparts, so long as much of the spectrum does not contain
energy~\cite{TroWakDua::2006::Random-filters,Rom::2008::Compressive-sensing,MisEld::2009::From-theory,TroLasDua::2009::Beyond-Nyquist:}.
Indeed, it has been shown that entire receiver chains for wideband
spectrum monitoring can be built around these new
devices~\cite{DavSchSla::2010::A-wideband-compressive}.

An implicit premise of DSA is that the spectrum consists mostly of
empty channels~\cite{bacchus:dyspan:2008,McHenry05a}. Thus, CS-based
ADCs offer an advantage to traditional sampling systems by reducing
the required sampling rate to represent the same spectrum. Furthermore, since CS
measurements can sometimes be processed directly without
algorithmic signal recovery, the reduction in acquired samples may also offer a potential reduction in the complexity of signal detection. These two properties of CS for spectrum sensing suggest a
new receiver design for DSA radios that offers the ability to sense
and detect signals with a reduction in computation and power
requirements.

The use of CS for DSA networks is not new and has been previously proposed~\cite{4632227,MenYinLi::2010::Collaborative-spectrum,4960089,tian:2007:cswavelets}.  However, the majority of this literature considers basic channel assumptions for both simulations and theory.  Furthermore, previous work primarily relies on non-linear signal reconstruction algorithms which may be a burden on speed and power, perhaps compounding issues in the second bottleneck above.  

In this paper, we explore the practical costs and benefits of CS for
DSA networks.  First, we review several fast and practical techniques
for energy detection without full reconstruction and provide
theoretical guarantees. We also define practical metrics to measure
the performance of these techniques. Second, we perform comprehensive
experiments comparing the techniques on a sparse public safety band
captured over the air.

Our experiments and analysis indicate both full and partial CS
reconstructions are relatively effective at detecting and avoiding
interferers.  At low compression ratios, the full reconstructions are
more accurate; however, under higher compression, the partial CS
reconstructions are both faster and more accurate.

The results of our work also provoke several important questions on the use of CS for DSA.  In what regimes is CS the most practical option for DSA acquisition and processing?  There is a more fundamental question: as DSA radios enable more use of the spectrum, the spectrum becomes less sparse.  How efficiently can we use the spectrum before CS becomes a victim of its own success? These are
questions that we will help to address in this paper through our
measurement data and results, and we revisit it in the conclusions.

The paper is organized as follows. Section \ref{sec:background}
provides a background on spectrum sensing and the theoretical underpinnings of compressive sensing. Section
\ref{sec:methods} discusses techniques that speed up DSA processing from CS measurements. Section \ref{sec:testbed} provides an empirical study of these techniques on data. In Section
\ref{sec:conclusion} we provide a general discussion on some of the practical
implications of this technology and what questions need to be
addressed in the future.

\section{Background}
\label{sec:background}

\subsection{Dynamic spectrum access}
\label{sec:dsa}

DSA relies on either  spectrum sensing or geolocation database querying techniques to determine which channels are free for transmission.
%
In response to industry demand, 
the Federal Communications Commission (FCC) moved to focus more on geolocation-based approaches  in 2010~\cite{FCCReconsiderationOrder}. Shortly afterwards, the UK communications regulator Ofcom released a consultation document regarding the use of geolocation~\cite{OfcomConsultation},

Despite these moves, however, spectrum sensing is still relevant and has a key role to play.  Specifically, geolocation-based techniques do not handle the case where multiple DSA radios exist in the same physical location (and operate over the same band).  Furthermore, the adoption of these standards does not guarantee that such databases are accurate or current.  
Thus, several international activities are currently focusing on spectrum sensing and characterization, including: IEEE 802.22~\cite{IEEE802.22}, 
IEEE SCC41~\cite{IEEESCC41}, and SE43~\cite{SE43}. 
International collaborative working groups, with a primarily European-based mandate, 
include COST-TERRA~\cite{COST-TERRA} and COST-IC0902~\cite{COST-IC0902}. 
European Union (EU) Seventh Framework Programme (FP7) projects include 
CREW~\cite{CREW}, CogEU~\cite{COGEU}, Acropolis~\cite{ACROPOLIS}, and FARAMIR~\cite{FARAMIR}.

\subsection{Spectrum sensing}
\label{sec:sensing}

Basic spectrum sensing is typically performed as follows. We acquire the spectrum with a receiver that is designed to some specified sensitivity (e.g., -114~dBm for TV whitespaces). Next, the noise floor is calculated and a threshold selected. Any received signal in excess of that threshold is deemed to be an occupied frequency; otherwise it is considered to be an available frequency. This threshold-based method is the basis of one of the first productized DSA radios from the DARPA XG project~\cite{mchenry:xg:2007}. Further studies have improved on how this information is used to make more informed decisions on what spectrum is usable~\cite{marshall:dyspan:2008}.

Techniques that improve upon basic energy detection have also been developed.  One such example is cyclostationary feature detectors~\cite{ge:2008:dyspan,Sutton2008}. Such detectors attempt to exploit the inherent cyclostationary features of man-made signals to distinguish them from noise. These attempts can often detect signals at lower noise levels than energy thresholding at the expense of additional sampling and processing time.

The detection process must work in conjunction with the RF front-end.  Once the signals are detected, the next layer up requires protocols to direct the behavior of a network in order to properly control a system of DSA radios~\cite{Buddhikot05}.

In this paper, we focus our attention on the detection of occupied
frequency bands over time.  We begin by considering the spectrogram of
a length-$W$ complex-valued signal $x$ via the short-time Fourier transform.
Specifically, we divide $x$ into a collection of $L$ blocks, each of
length $N$ (so $W=NL$).  Let $x^l$ be the $l$-th block of data and let
$\nu^l$ be the DFT of $x^l$.

Thus, we can write our $N\times L$ spectrogram as
\begin{equation}
\mathcal{S}_{N,L}(x) = \left|[\nu^{1}, \ldots, \nu^{L}]\right|^{2},
\end{equation}
where $|\cdot|^{2}$ is applied element-wise.  

We now group $\beta$ consecutive frequencies into $B=N/\beta$ channels
and $\gamma$ consecutive blocks into $G=L/\gamma$ time slots.

We imagine that an interferer is active on a channel $b$ and a time
slot $g$.  Given a channel $b$ and a time slot $g$, the pair $(b,g)$
corresponds to a collection of frequencies and blocks (i.e., a
rectangle within the spectrogram).  The power of the pair $(b,g)$ is
the sum of the power of the constituent (frequency, block) pairs in
the spectrogram.

Given a threshold $\theta$, we say that a pair $(b,g)$ contains an
interferer if 
\[\mbox{Power }(b,g) \geq \theta\]
Our goal is to determine precisely which $(b,g)$ pairs are
interferers, so that we can avoid them.

The spectrogram model as described above assumes that we have access
to direct Nyquist samples of a wide bandwidth of data.  As discussed
earlier there are practical limitations to acquiring these samples.
Thus, we now turn to compressive sensing for more efficient
acquisition.

\subsection{Compressive sensing}
\label{sec:cs}

Since $x$ is processed in blocks of length $N$, for the remainder of this section $x$ will refer to any subsection of the full signal.  In the compressive sensing framework~\cite{Can::2006::Compressive-sampling,Don::2006::Compressed-sensing,CanRomTao::2006::Stable-signal},  linear measurements of a discrete vector $x\in \mathbb{C}^{N}$ are computed as
\begin{equation}
\label{eq:mmodel}
y = \Phi x,
\end{equation}
where $\Phi$ is an $M\times N$ measurement matrix with $M \ll N$ rows and $y$ is the vector of $M$ measurements.  In analog systems, $x$ can be thought of as the Nyquist-rate (or above) samples of an analog signal $x(t)$ and a hardware measurement system corresponds to an analog operator $\bar{\Phi}$ that produces discrete measurements such that $y = \Phi x = \bar{\Phi}(x(t))$. For the remainder of this paper we consider the discrete formulation.

The measurement model (\ref{eq:mmodel}) is underdetermined and thus not invertible in general. However, a key insight in CS is that by restricting $x$ to a smaller class of signals, we may be able to invert the system.  Specifically, we consider $x$ that are $K$-sparse, i.e., signals that have only $K\lll N$ nonzero values.  To see why this signal model works, suppose we knew the positions of these nonzero coefficients and $K<M$, then we could obtain an overdetermined, invertible submatrix of $\Phi$ consisting of the columns corresponding to the support of $x$, and use this to recover $x$ from $y$.

An important feature of CS systems is that $x$ can be sparse in an arbitrary transform basis, such as the DFT-basis or the wavelet basis.  Specifically, the model allows signals of the form $x = \Psi \nu$ where $\Psi \in \mathbb{C}^{N\times N}$ is an orthonormal basis and $\nu \in \mathbb{C}^{N}$ is a sparse vector.  Furthermore, CS allows for \emph{compressible} signals, i.e., those that can be closely approximated by a $K$-sparse signal, such as $K$-sparse signals with noise, or signals whose sorted coefficient magnitudes decay rapidly with some power law. 

Given the signal model, we can write $y = \Phi x = \Phi\Psi \nu = A\nu$.  One desirable property of $A$ would be that for all $K$-sparse signals, we obtain a unique $y$.  Indeed, this idea is similar to the so-called \emph{restricted isometry property} (RIP)~\cite{CandesDLP}:
\begin{equation}
\label{eq:RIP}
(1-\delta_{K})\|\nu\|_{2}^{2} \leq \|A\nu\|_{2}^{2} \leq (1+\delta_{K})\|\nu\|_{2}^{2},
\end{equation}
for all $K$-sparse $\nu$ and some matrix-dependent constant $\delta_{K}$.   Supposing that $A$ satisfies the RIP, then the program \emph{Basis Pursuit}:
\begin{equation}
\widehat{\nu} ~~\gets~~ \min_{\nu}\|\nu\|_{1}~\mathrm{s.t.}~y = A\nu,
\end{equation}
will recover $\nu$ exactly.  Furthermore, if the measurements contain noise $y = A\nu + n$ with $\|n\|_{2}<\epsilon$, then the program \emph{Basis Pursuit Denoising} (BPDN):
\begin{equation}
\label{eq:BPDN}
\widehat{\nu} ~~\gets~~  \min_{\nu}\|\nu\|_{1}~\mathrm{s.t.}~\|y - A\nu\|_{2} < \epsilon,
\end{equation}
will recover $\nu$ with error
\begin{equation}
\|\nu - \widehat{\nu}\|_{2} \leq C_{1}\epsilon + C_{2}\frac{\|\nu- \nu_{K}\|_{1}}{\sqrt{K}},
\end{equation}
where $C_{1}$ and $C_{2}$ are constants and $\nu_{K}$ denotes the best $K$ term approximation of $\nu$~\cite{CanRomTao::2006::Stable-signal}.  It has also been demonstrated that several other greedy and iterative algorithms will solve this problem with similar guarantees as long as $A$ has the RIP~\cite{cosamp,BluDav::2008::Iterative-hard}.  While precise values of $C_{1}$ and $C_{2}$ are unknown, empirical evidence suggests that these constants can be smaller than $2$ (and can be close to 1).  For the case of additive noise, this implies that we take a $6$~dB hit in SNR during reconstruction. 

Remarkably, it has been shown that a large class of matrices satisfy the RIP.  If $M = O(K\log(N/K))$, then matrices whose elements are drawn from a sub-Gaussian distribution\footnote{A sub-Gaussian distribution is a distribution whose moment generating function is bounded by that of the Gaussian.} will satisfy the RIP with high probability~\cite{Dav::2010::Random-observations}.  Thus, our system satisfies $K < M \ll N$ and yet we can achieve robust signal recovery.  

\subsection{The power benefits of reducing the number of samples}
A primary factor in ADC power consumption is quantization.  Specifically, an ADC is composed
of two main parts: a ``sample and hold'' operation to discretize in time followed by ``quantization'' to discretize real-valued samples.  It is well understood that the majority of the ADC's power is consumed in the quantizer~\cite{le_adc2005}. By reducing the number of measurements we acquire,  we require fewer quantization operations, and thus can expect significantly greater power savings. 

Furthermore, in some systems acquisition of fewer samples will correspond to a lower ADC sample-rates.  Reducing the sampling rate of an ADC enables significant power savings. Specifically, it has been empirically demonstrated (via a survey of commercial devices) that power consumed by an ADC grows at a rate of $1.1 \times f_s$ where $f_s$ is the converter's sampling rate~\cite{le_adc2005}. For example, an 8-bit flash ADC at 200~Msps consumes 2320~mW of power (or 11.6~nJ/sample), but an 8-bit flash ADC at 20~Msps only
consumes 150~mW (or 7.5~nJ/sample). So by reducing the sampling rate
by a factor of 12.5, we have also reduced the power consumption by a
factor of roughly 15.5.   In the next subsection, we will see that quantization also plays a role in the maximum achievable sample rate, and how CS can be used to mitigate this issue.

\subsection{CS in practice}
Several practical analog-to-digital converters have been proposed for use within the CS framework~\cite{TroWakDua::2006::Random-filters,Rom::2008::Compressive-sensing,MisEld::2009::From-theory,TroLasDua::2009::Beyond-Nyquist:,SlaLasDav::2011::The-compressive-multiplexer}.  The matrix representation of many of these devices, while significantly more structured than a random matrix, have also been shown to satisfy the RIP.  Additionally, many forms of $A$ and $A^{T}$ can be efficiently computed using fast transforms, rather than matrix multiplication.

As a brief example, consider the compressive multiplexer (CMUX)~\cite{SlaLasDav::2011::The-compressive-multiplexer}.  This device demodulates $J$ spectrum channels to baseband, modulates them by a pseudo-random $\pm 1$ chipping code (operating at the Nyquist rate of the channels),  sums the channels together, and samples the output with a single (off-the-shelf) ADC operating at the rate of one channel.  The resulting CS matrix consists of $J$ diagonal submatrices appended together, each submatrix corresponding to a channel and containing the $\pm1$ sequence used on that channel.  Beyond reducing the total number of samples acquired, in this architecture CS enables an ADC that is sub-Nyquist over the combined bandwidth of the channels.

An alternative example of how this framework can be applied is by separating the ``sample and hold'' stage of the ADC from the ``quantization'' stage.  Specifically, Lyric Semiconductor has fabricated a chip that can perform certain matrix multiplications in the analog domain.  Such a device allows us to discretize and input signal in time and perform CS projections without quantizing.  Once the projections are computed, we only need to quantize $M \ll N$ values.  This permits a significant increase in overall sample rate (or maximum bandwidth acquired) since quantizers are not just the most power hungry components but also the primary bottleneck in ADC speeds~\cite{le_adc2005}.

In some cases we may wish to perform techniques such as detection without fully recovering the signal before processing.  It has been demonstrated that detection and classification can be performed directly on the measurements $y$~\cite{DavDuaWak::2007::The-smashed-filter}.  Similarly, filtering and even FM-demodulation do not require signal recovery before processing~\cite{DavSchSla::2010::A-wideband-compressive}.  This leads to significant speedups in processing time since the reconstruction procedure is omitted and only $M$ measurements are processed rather than the large number of samples $N$.  We will attempt to harness this capability where possible in this paper.

\section{Practical Methods for DSA from CS Measurements}
\label{sec:methods} 

\subsection{Reconstruction-based methods}
The CS framework described above can be applied to the DSA problem in
a straightforward fashion: we simply perform signal recovery via CS
(\ref{eq:BPDN}) and then use the reconstructions to estimate the
presence of interferers.  However, this approach has several
drawbacks, the primary being the speed of the reconstruction
algorithm.  Several relatively fast algorithms have been developed,
but in real-time applications even these fast reconstructions may
prove to be too slow.  The standard method of energy detection in a
DSA system only involves performing FFTs, a power calculation, and
thresholding.  All of these are computationally inexpensive,
especially compared to the optimization routines used in CS signal
reconstruction.


While CS theory tells us that the reconstructed signal will be close to the original, in spectrum sensing, we are not interested in the signal itself, but the energy in different spectrum channels. We now discuss some techniques that avoid full reconstruction, yet have theoretical guarantees on the energy estimate for specific frequencies or channels.  

\subsection{Frequency testing}
We can compute a rough estimate of the DFT coefficients of the signal by
\begin{equation}
\label{eq:rough}
\widehat{\nu} \gets A^{T}y,
\end{equation}
where $A = \Phi \Psi$ and $\Psi$ is the DFT basis.  We will denote this method as the ``transposition technique'' for the remainder of the paper.  Applying detection techniques after this operation can be thought of as a special case of the \emph{smashed filter}~\cite{DavDuaWak::2007::The-smashed-filter}.  We refer the reader to theoretical guarantees on detection and classification in CS therein.

As mentioned earlier, in many practical CS systems, the operation $A^{T}$ can be computed quickly using transform-based techniques.  Indeed, even if $A$ is treated as a matrix, the computation required for any optimization or iterative algorithm will be greater than for (\ref{eq:rough}).

Supposing that $A$ satisfies the RIP of order $2K$ with constant $\delta_{2K}$, then for any two $K$-sparse signals $\nu, \omega \in \mathbb{R}^{N}$ it is shown in~\cite{Can::2008::The-restricted-isometry} that
\begin{equation}
|\langle A\omega, A\nu \rangle - \langle  \omega, \nu \rangle| \leq \delta_{2K} \|\omega\|_{2}\|\nu\|_{2}.
\end{equation}
We can use this to bound the accuracy of each DFT coefficient estimate $\widehat{\nu}_{i} = \langle A_{i}, y \rangle$, where $A_{i}$ is the $i$-th column of $A$.  Let $\omega = e_{i}$, the $i$-th canonical basis vector containing all zeros except for index $i$ which contains the value $1$.  Then $A_{i} = A e_{i}$ and $\nu_{i} = \langle e_{i},  \nu \rangle$, yielding
\begin{equation}
|\langle A_{i}, y \rangle - \nu_{i}| \leq \delta_{2K}\|\nu\|_{2}.
\end{equation}
Thus, we have that the difference between $\nu_{i}$ and its estimate $\widehat{\nu}_{i}$ is bounded from both directions.  We note that (\ref{eq:rough}) simply computes all of these coefficient estimates.  Interestingly, the maximum error on each coefficient estimate will be proportional to the norm of the entire signal $\nu$.

While we desire small $\delta$ for our CS systems, this constant is elusive in practice. 

\subsection{Channel testing}
\label{sec:methods:channels} 
Rather than computing individual frequencies, we may be interested in testing the energy in several spectrum segments.  Specifically, let $\Lambda_{b} \subset \{1,\cdots,N\}$ be a coefficient index set of cardinality $|\Lambda_{b}|$ corresponding to the DFT coefficients in frequency band $b$.  In this scheme, we would test each band by computing
\begin{equation}
\label{eq:bintest}
h_{b} = \|(A^{T}y)_{\Lambda_{b}}\|_{2}^{2}.
\end{equation}
By combining proposition 3.1 and corollary 3.3 of \cite{cosamp}, we find that for $S \geq |\Lambda_{b} \bigcup \mathrm{supp}(\nu)|$ and $A$ satisfying the RIP, 
\begin{equation}
(1-\delta_{S})\|\nu_{\Lambda_{b}}\|_{2}^{2} \leq h_{b }\leq (1+\delta_{S})\|\nu_{\Lambda_{b}}\|_{2}^{2} + \delta_{K}\|\nu_{\Lambda_{b}^{c}}\|_{2}^{2},
\end{equation}
where $\Lambda_{b}^{c}$ is the compliment index set of $\Lambda_{b}$ and $\mathrm{supp}(\nu)$ is the support of $\nu$.  This implies that if $A$ satisfies the RIP of order $S$ (with small $\delta_{S}$), then the frequency bin energy test (\ref{eq:bintest}) should be close to the original energy in that frequency bin.  

As was the case with frequency testing, we do not know $\delta$.  Therefore our experiments will play a key role in understanding the practical performance of this method.

\section{Empirical Study}
\label{sec:testbed}

\subsection{Spectrum data sources}
\label{subsec:spectrumdata}

\begin{figure}[htbp] 
   \centering
   \includegraphics[width=1.0\linewidth]{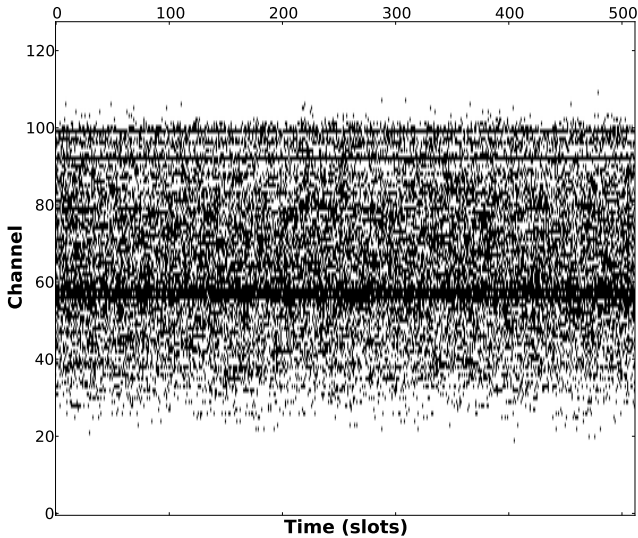} 
   \caption{A spectrogram sample of the spectrum captured around 749~MHz. These signals were sampled at 200~MS/sec.}
   \label{fig:ws749_spec}
\end{figure}

One of the key challenges of DSA system concepts is converting theory to practical implementation and characterizing performance in real-life scenarios. To this end, we applied CS for signal detection using live, over the air, spectrum captures. Sampled IQ samples from spectrum segments in the UHF band were captured using an Anritsu MX2690A spectrum analyzer sampling at 200\,MS/sec. Each spectrum capture spanned 100\,MHz of activity over a 0.5 second duration in Dublin city center, Ireland, during July, 2010. The center frequencies chosen for our study were 220.5\,MHz, 450\,MHz and 749\,MHz. These were selected to represent a cross section of spectrum activity covering private mobile radio (PMR), emergency services, and amateur radio use in addition to digital audio broadcasting, and analog and digital terrestrial television. An example of the spectrogram of the data collected at a center frequency of 749~MHz is shown in Figure~\ref{fig:ws749_spec}.

These frequencies were selected based on a few criteria. They are in
the range of digital dividend and TV white space frequencies being actively discussed for DSA use and they
tend to feature a wide variety of active users and services. 

The experiments and analysis in this paper are all based on live
captures. For the purposes of the experiment, we collected our data by
using a traditional Nyquist sampling receiver and performed the
compressive sampling on these digitized samples. As we have discussed
in the introduction, in a fielded system we imagine that the
compressive operation would be performed by, for instance, an analog
DFT.  In this way, we provide a bridge from simulated CS
experiments used in the cited works to full CS spectrum sampling
enabled by these new devices. We show through our experiments
that CS indeed works on live signals.

\subsection{CS measurement system}
In our experiments, we implemented a computationally efficient CS
measurement system.  Specifically, we partition the signal into blocks
of length $J=1024$ complex samples.  On each block we pseudo-randomly
permute the signal entries, compute a DFT, and then retain a random
subset of coefficients from the DFT.  We retain a total of $M$
measurements.  This results in a block-diagonal $M\times N$ matrix
$\Phi$ with each block being a subset of DFT rows with permuted
columns.  For the sparsity basis, we choose the $N\times N$ DFT
matrix.  Because $A=\Phi\Psi$ and $A^T$ are composed from 
permutations and DFTs, they can be computed efficiently.

We note that the system $\Phi$ is practical in that it can be computed by the Lyric chip described earlier.  However, since the main focus of this paper is on reconstruction speed and detection accuracy, we simulated the operation $\Phi$ in software.


\subsection{Experimental setup}
\label{subsec:setup}

\begin{figure}[htbp] 
   \centering
   \includegraphics[width=1.0\linewidth]{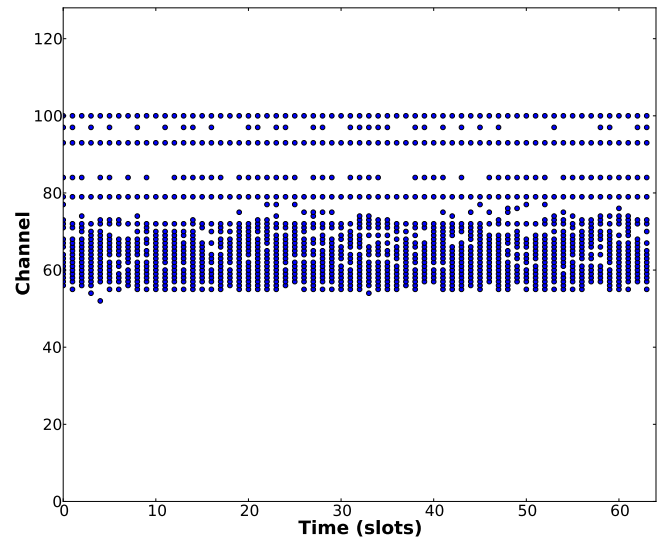} 
   \caption{Channels and time slots that exceed a power threshold.  We
     define these regions as interferers.}
   \label{fig:example0}
\end{figure}

The experiments proceed as follows. We use the notation from
Section~\ref{sec:background}.  We divide the signal $x$ into blocks of
length $J=1024$ complex samples.  We group $\gamma=64$ blocks into a
single time slot.  Our channels are 8 frequencies tall, so there are
$1024/8=128$ channels.  If we consider the corresponding spectrogram (with channels instead of frequencies), a stereotypical pattern of interferers is illustrated in Figure~\ref{fig:example0}.

Figure~\ref{fig:example0} was created using the power thresholding
method described Section~\ref{sec:sensing} on the signal shown in
Figure~\ref{fig:ws749_spec}.

For each block, we draw a new measurement system $\Phi$ of size
$M\times N$ as described above and reconstruct using both BPDN (via SPGL1~\cite{BergFriedlander:2008,spgl1:2007}) and the transpose method to estimate power.  Reconstruction
is performed on each disjoint block of $x$ and then transformed back
into the time domain to produce a final estimate.  We then perform the
DSA procedure proposed in Section~\ref{sec:methods}.  This procedure
is repeated for various compression ratios $M/N$.

\subsection{Metrics and results}
\label{subsec:results}

Before we launch into the results, we need a brief digression on
metrics of performance.  Because our results are based on real world
data, we need to define what we mean by an ``interferer''.  As we
discussed in Section~\ref{sec:background}, we choose a constant
$\theta$ and define any channel and time slot that exceeds power
$\theta$ as an interferer.  This method provides us with an estimated
ground truth because, as a live signal capture, we do not know the
exact nature of all of the signals included in the capture.

However, when we estimate the power in each channel and
time slot with some reconstruction algorithm, we can use a separate
power threshold $\theta'$ and avoid all (channel, time slot) pairs
that exceed power $\theta'$.  If we wish to avoid many interferers, we
can use a small value for $\theta'$.  However, a small $\theta'$ will
cause us to discard innocuous (channel, time slot) pairs.  Therefore,
there is a tradeoff between detecting true interferers and wasting
unallocated spectrum. 

We begin by examining the effect of the compression ratio $M/N$.
Suppose we reconstruct the signal and choose our threshold $\theta'$
so that we can detect 90\% of the interferers.  We then avoid certain
(channel, time slot) pairs because they appear to have larger power
(according to our reconstruction).  The more accurate our
reconstruction, the fewer pairs are avoided; in an optimal
reconstruction, we would avoid exactly 90\% of the true interferers.

\begin{figure}[htbp] 
   \centering
   \includegraphics[width=1.0\linewidth]{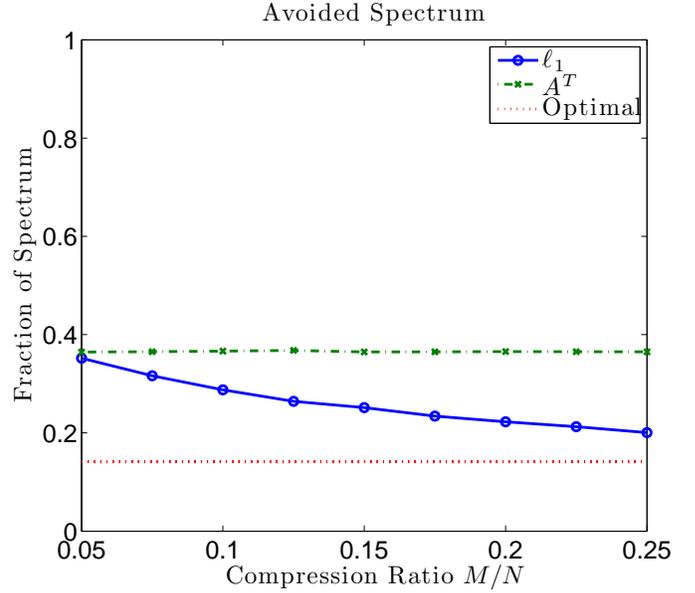} 
   \caption{ Each reconstruction technique produces an estimate of the
     power per channel and per window of time.  If we set our power
     threshold so that we correctly detect 90\% of the interferers, we
     waste a certain fraction of the spectrum.  This wasted fraction
     is illustrated above.  }
   \label{fig:gg1}
\vspace{.3in}
\end{figure}

As the compression ratio $M/N$ increases, we preserve more data, so
our reconstruction should become more accurate and the fraction of
avoided (channel, time slot) pairs should decrease.  We illustrate
this effect in Figure~\ref{fig:gg1}.  Notice that increasing the
compression ratio helps the $\ell_1$ reconstruction, but does not
markedly improve the transposition technique.

\begin{figure}[htbp] 
   \centering
   \includegraphics[width=.8\linewidth]{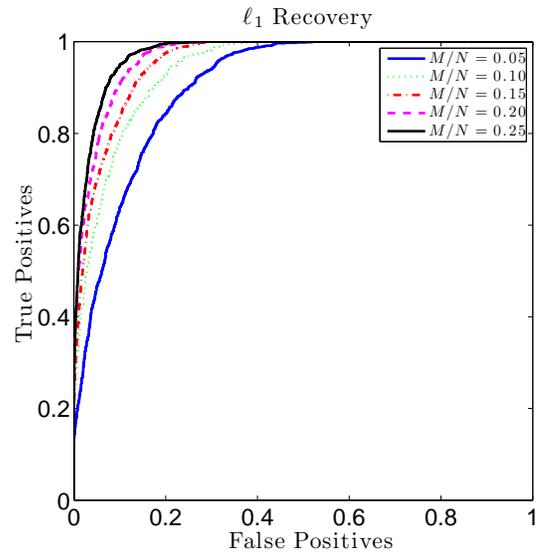} 
   \caption{ROC curves for $\ell_1$ reconstruction at various
    compression ratios.}
\label{fig:rocL1}
\vspace{.3in}
\end{figure}

\begin{figure}[htbp] 
   \centering
   \includegraphics[width=.8\linewidth]{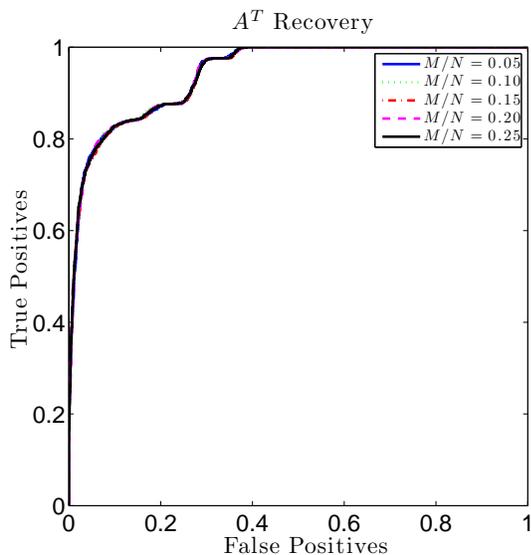} 
   \caption{ROC curves for $A^T$ reconstruction at various
    compression ratios.}
 \label{fig:rocTr}
\vspace{.3in}
\end{figure}

As we decrease the $\theta'$ parameter in the reconstruction, we can
increase the number of interferers that we detect (true positives),
but at the cost of occasionally increasing the number of
non-interferers that we mistakenly label as interferers (false
positives).  We can construct an ROC curve to summarize this
trade-off.  In Figure~\ref{fig:rocL1}, we plot a series of ROC curves
corresponding to different compression ratios $M/N$ from an $\ell_1$
reconstruction.  In Figure~\ref{fig:rocTr}, we construct the same
graph for the transposition technique.

As hinted by Figure~\ref{fig:gg1}, the transposition technique is
relatively insensitive to the compression ratio--increasing the ratio
does not improve the recovery significantly.  With the transposition
technique, the curves in Figure~\ref{fig:rocTr} for different
compression ratios are so similar that they are hard to distinguish
visually.  The $\ell_1$ technique exploits the reduce compression more
effectively, so the ROC curves in Figure~\ref{fig:rocL1} improve in a
more obvious fashion.

\begin{figure}[htbp] 
   \centering
   \includegraphics[width=.8\linewidth]{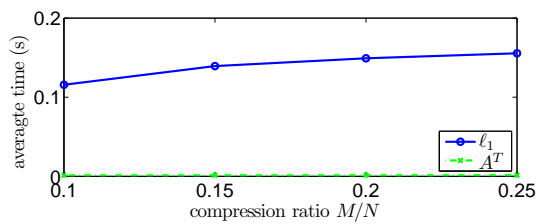} 
   \caption{The relative speed of the full $\ell_1$ reconstruction
     versus the transposition technique.}
  \label{fig:example2}
\vspace{.3in}
\end{figure}

Finally, we can compare the running time of the two algorithms.
As Figure~\ref{fig:example2} illustrates, the transposition technique
is significantly faster than than the $\ell_1$ recovery technique.

\section{Discussion}
\label{sec:conclusion}

In this paper, we have presented results that suggest compressive
sensing is a viable method for enhancing spectrum sensing for DSA
systems. Because CS allows a front-end to collect many fewer samples
than conventional ADCs, a CS-based DSA radio could sense more
spectrum with the same sampling requirements or the same spectrum with
reduced sampling requirements, resulting in cheaper and more
power-efficient systems. We have demonstrated that CS can represent
signals present in traditional whitespace spectrum, making it possible
to then detect these signals for DSA.



In particular, we reviewed several fast and practical techniques for energy detection without full reconstruction and explored the practical costs and benefits of CS for DSA networks. As part of our analyses, we performed comprehensive experiments comparing the techniques on real signals captured over the air. 

This study has raised a number of follow-on questions. While
improvements in digital processors will continue to bolster the use of
CS in real-time embedded radios, another continuous growth factor will
have the opposite effect: spectrum occupancy. CS works because of the
knowledge that the sample set is sparse. Current spectrum sparsity
that DSA radios hope to fill will negatively impact the usefulness of
compressive sensing. The irony of this approach could be that it
becomes a victim of its own success. More work remains to be done on
this problem to see the real potential of CS for DSA.

Another question that we did not address in this paper is the use of
various thresholds for detecting different signals. With various
coexisting heterogeneous radios in the same sampled spectrum,
different levels of detection may actually be required. While we did
not experiment with this concept directly, we believe that the
extension of these algorithms to such a heterogeneous detection
environment is possible. Indeed, \ref{sec:methods:channels} begins
setting up just such a capability.

Compressive sensing is a fairly new analytical tool for use in many
domains, including DSA and this and other papers have
demonstrated. Still, there are many issues yet to be addressed and
proven for their effective and efficient use in spectrum sensing and
signal detection.

\bibliographystyle{IEEEtran}
{\small \bibliography{dyspan}}

\end{document}